\documentclass{aa} 
\usepackage{graphicx}

\authorrunning{C. Li}
\titlerunning{Acceleration characteristics of SEPs on 14/07/00}

\begin{document}
   \title{The acceleration characteristics of solar energetic
   particles in the 2000 July 14 event}

   \author{C. Li \inst{1}, Y. H. Tang \inst{1}, Y. Dai \inst{1},
   W. G. Zong \inst{2}, \and C. Fang \inst{1}}


   \institute{Department of Astronomy, Nanjing University, Nanjing 210093,
   China\\ \email{yhtang@nju.edu.cn, lic@nju.edu.cn}
   \and
   National Center for Space Weather Monitoring and Warning,
   Meteorological Administration, Beijing 100081, China\\
   \email{wgzong@nsmc.cma.gov.cn}}

   \date{Received 6 June 2006 / Accepted 4 August 2006}

   \abstract
  {}
  {In large gradual solar energetic particle (SEP) events, especially
the ground-level enhancement (GLE) events, where and how energetic
particles are accelerated is still a problem.}
   {By using imaging data from TRACE, Yohkoh/HXT, SOHO/MDI and SOHO/EIT,
along with the data from the GOES, Apatity NM, and SOHO/LASCO CME
catalog, the evolution of the X5.7 two-ribbon flare and the
associated SEP event on 14 July 2000 are studied.}
   {It is found that the magnetic reconnection in this
event consists of two parts, and the induced electric field
$E_{rec}$ is temporally correlated with the evolution of hard
X-ray and $\gamma$-ray emission. In particular, the first hard
X-ray and $\gamma$-ray emission peak occurred at 10:22 UT,
corresponding to the magnetic reconnection in the western part of
the flare ribbons and the maximum $E_{rec}$ of $\sim$ 9.5 V/cm;
the second emission peak at 10:27 UT, corresponding to the eastern
part and the maximum $E_{rec}$ of $\sim$ 13.0 V/cm. We also
analyze the SEP injection profiles as functions of time and
CME-height, and find two-component injection which may result from
different acceleration mechanisms.}
   {A reasonable conclusion is that
reconnection electric field makes a crucial contribution to the
acceleration of relativistic particles and the impulsive component
of the large gradual SEP event, while CME-driven shocks play a
dominant role in the gradual component.}


   \maketitle

\section{Introduction}
Solar energetic particles are one manifestation of violent energy
releases on the Sun. It is widely accepted that SEP events can be
divided into two classes: impulsive and gradual (Cane et al.
\cite{Cane86}; Kallenrode et al. \cite{Kallenrode92}). The
impulsive SEPs are accelerated in flares, while the gradual ones
are accelerated by CME-driven shocks (Reames \cite{Reames90};
Reames \cite{Reames99}; Kahler \cite{Kahler01}; Kallenrode
\cite{Kallenrode03}). ``Mixed" events consisting of both particles
accelerated in flares and at the shocks were considered to be
nonexistent (Reames \cite{Reames02}). In particular, Kahler
(\cite{Kahler94}) examined the SEP injection profiles of three
ground-level enhancement (GLE) events as the function of the
height of the associated coronal mass ejections (CMEs). They
suggested that in these events SEP injection appears to only
result from a single CME-driven shock and not from the flare
impulsive phase.

In 1992, Kallenrode et al. found the most common energetic proton
events are those exhibiting both an impulsive phase and shock
emission, the ``pure gradual" events are relatively rare. They
suggest that between these two classes there should be ``mixed" or
``hybrid" events with both impulsive and shock acceleration.
Cliver et al. (\cite{Cliver95}) then expanded the two-type
classification to include the ``mixed" events, in which the SEP
event contains a mixture of flare-accelerated and
CME/shock-accelerated particles. Recent observations and studies
provide more and more evidence of the existence of mixed SEP
events. Cohen et al. (\cite{Cohen99}) found that the charge states
of 12 elements with energies of 12 -- 60 MeV/nucl and source
temperatures of (3 -- 6)$\times 10^{6}$ K, are significantly
higher than at lower energies. It rules out acceleration out of
the ambient material (Kallenrode \cite{Kallenrode03}). In
addition, Cane et al. (\cite{Cane03}) find that in some intense
SEP events, the time-intensity profiles exhibit two peaks: the
earlier one with a high Fe/O ratio, the later one a low Fe/O
ratio. A much simpler interpretation is that the earlier one is
flare-related and the later one is CME-driven shock related. Dai
et al. (\cite{Dai05}) find that some large gradual SEP events have
an initial impulsive component, which may stem from flare
acceleration. Shocks could also add substantial protons and
contribute to the gradual component (Cane et al. \cite{Cane86}).
And a recent simulation of mixed particle acceleration by Li \&
Zank (\cite{Li05}) indicated that the SEP time-intensity profile
has an initial rapid increase followed by a plateau similar to a
pure shock case.

An outstanding solar event occurred on 14 July 2000, ``Bastille
Day'', comprising an X-class flare, a fast halo CME, and a large
particle intensity enhancement in interplanetary space. It offers
an excellent opportunity to extend our knowledge about the
acceleration and transport process of solar energetic particles.
The GOES X5.7/3B two-ribbon flare, which occurred in the NOAA
region 9077 (N22W07) at 10:03 UT, has been thoroughly studied by
Aschwanden \& Alexander (\cite{Aschwanden01}), Fletcher \& Hudson
(\cite{Fletcher01}), and Masuda et al. (\cite{Masuda01}). Their
studies mainly focus on the generation of EUV and hard X-ray
sources and on the corresponding magnetic structure. One of their
important results is that flare ribbons map the footprints of
magnetic field lines newly linked by reconnection, along which
nonthermal particles bombard the lower atmosphere and lose their
energy.

In this paper, we go on to use high-cadence observations from the
Transition Region and Coronal Explorer (TRACE), along with the
longitudinal magnetic field from the Michelson Doppler Imager
(MDI) on board SOHO, to evaluate the magnetic reconnection
electric field. This enables us to estimate a charged particle's
energy gain in the active region. Because this event was
associated with a fast halo CME, we also analyzed the SEP
injection profiles as the functions of time and the CME-height. We
then discuss the roles of flare and CME-driven shock in producing
energetic particles. This paper is organized as follows: In Sect.
2, we analyze the reconnection electric field, comparing it with
hard X-ray emission, and investigate the SEP injection profiles.
Section 3 is dedicated to discussions, and conclusions are drawn
in Sect. 4.

\section{Data analysis and results}

\subsection{Reconnection electric field}
In the standard model of flare evolution, a rising reconnection
region results in ribbons moving apart through the photospheric
magnetic field. The separation motion of flare ribbons that sweep
through the magnetic field lines corresponds to the rate of
magnetic reconnection in the corona, where the reconnection
current sheet (RCS) is generated (Forbes \& Priest
\cite{Forbes84}, Forbes \& Lin \cite{Forbes00}). We measured the
magnetic reconnection rate in the form of a reconnection electric
field. In a 2D-configuration approximation it can be given by
$E_{rec}=VB$ (Qiu et al. \cite{Qiu02}), where $V$ is the
separation velocity of flare ribbons and $B$ the magnetic field
that the ribbons sweep through. Since the flare occurred near the
disk center, $B$ can be approximately taken as the longtitudinal
component of the magnetic field obtained from the SOHO/MDI
magnetogram at 09:36:01 UT before the flare. The mean reconnection
electric field induced inside the current sheet is obtained by
averaging the electric field inferred in each of the two ribbons.
\begin{figure}
\centering
\includegraphics[width=8cm]{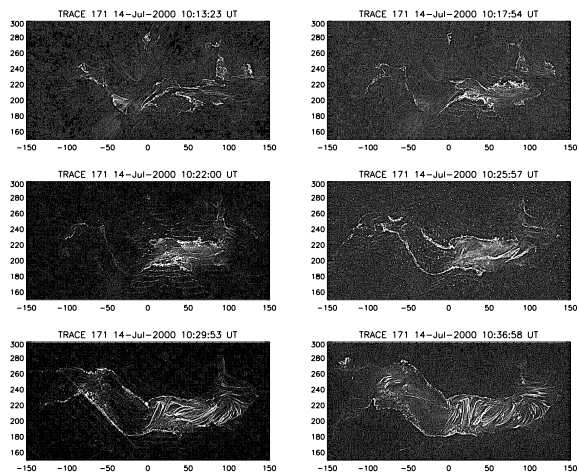}
\caption{High-pass filtered TRACE $171\rm \AA$ images at 6 times.
Note that flare ribbons separate first in the western part, and
later in eastern part.} \label{fig1}
\end{figure}

We used the high-cadence ($\sim1$ minute) 171$\rm\AA$ TRACE
observations covering the time interval of 10:11 UT -- 10:59 UT to
measure the ribbon separation of the flare. To enhance the spatial
structure of the flare ribbons, we applied a high-pass filter by
subtracting a smoothed image (smoothed by 3$\times$3 pixels) from
the original images. Figure 1 shows six of these images of the
active region at different times. It is clear that the separation
of the flare ribbons occurred first in the western part and then
in the eastern part.

To overlay the positions of the flare ribbons on the magnetogram,
we first resized the TRACE white light image (at 09:35:16 UT) to
the same pixel size as an MDI continuum image (at 09:35:31 UT),
and cross-correlated the two images. This procedure returns a
value for the offset between the TRACE and MDI images. With this
offset correction, the positions of the flare ribbons for each
TRACE 171$\rm\AA$ image can be traced and overlaid on the MDI
magnetogram. Since the time interval we studied is $\sim$ 50
minutes, the solar rotation must be taken into account. We also
reconstructed hard X-ray images (M2 channel) made during the two
peaks of the flare emission (10:22:45 -- 10:23:10 UT and 10:26:58
-- 10:27:04 UT), and overlaid them on the MDI magnetogram. Figure
2 shows the positions of EUV ribbons and hard X-ray sources
overlaid on the MDI magnetogram.
\begin{figure}
\centering
\includegraphics[width=8cm]{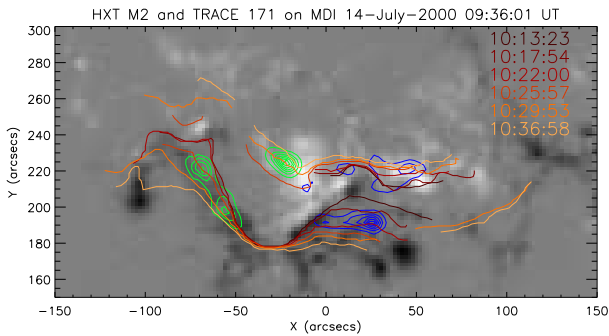}
\caption{Positions of the EUV ribbons and hard X-ray sources at
the two peaks of flare emission (Blue: 10:22:45 -- 10:23:10 UT and
Green: 10:26:58 -- 10:27:04 UT) overlaid on the magnetogram.
Different color lines indicate the EUV ribbons at different times,
and a white background indicates positive longitudinal magnetic
field, and black negative.} \label{fig2}
\end{figure}

As Fig. 1 shows, the eruption of the flare is divided into two
stages. The first eruption occurred in the western part of the
flare ribbons between 10:11 UT and 10:24 UT, and then it triggered
the second eruption at the eastern part after 10:24 UT. This
indicates that the reconnection current sheet should consist of
two parts, corresponding to the evolution of the two parts of the
flare ribbons. Comparing the electric field $E_{rec}$ inferred
from the two parts with flare nonthermal emission at hard X-ray,
we find good temporal correlation between them, as shown in Fig.
3. In particular, the first hard X-ray emission peaks at 10:22 UT,
corresponding to the magnetic reconnection in the western part of
the flare ribbons and the maximum $E_{rec}$ of $\sim$ 9.5 V/cm and
the second one occurred at 10:27 UT, corresponding to the magnetic
reconnection in the eastern part and the maximum $E_{rec}$ of
$\sim$ 13.0 V/cm. Two $\gamma$-ray emission peaks were also
reported at 10:22 UT and 10:27 UT (Share et al. \cite{Share01}).
\begin{figure}
\centering
\includegraphics[width=8cm]{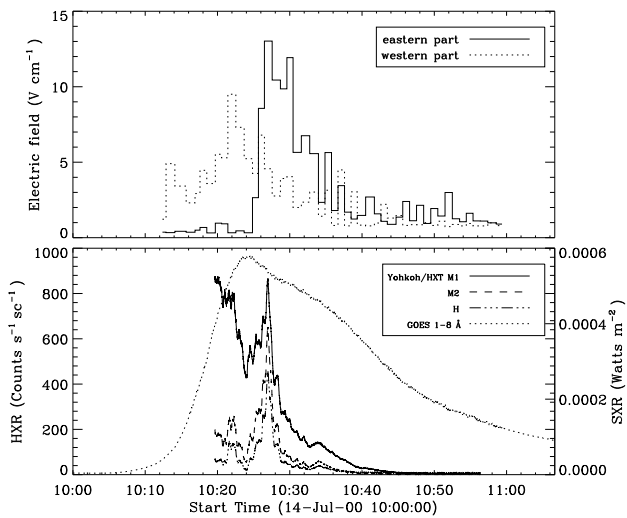}
\caption{Reconnection electric field $E_{rec}$ inferred from two
parts of the X5.7 flare ribbons compared with hard X-ray and soft
X-ray light curves. The three Yohkoh/HXT energy bands(M1, M2, H)
cover the energy ranges 23 -- 33, 33 -- 53 and 53 -- 93 keV,
respectively.} \label{fig3}
\end{figure}

From Fig. 2, we can also find that the hard X-ray sources, in
which energetic particles precipitated along magnetic field lines
newly linked by reconnection, are just located at the area where
the flare ribbons have the most rapid separation. The good
temporal and spatial correlation indicates a physical link between
magnetic reconnection and energy release in flares, and also
suggests that reconnection electric field $E_{rec}$ plays an
important role in accelerating nonthermal particles.

\subsection{SEP injection profiles}
The 2000 July 14 GLE event was accompanied by both an X5.7 flare
and a fast halo CME with a speed of 1674 km/s. To understand the
roles of the flare and the CME-driven shock in producing solar
energetic particles, both the proton and cosmic ray intensities
were compared to the CME height-time profile. The proton intensity
was obtained from Geostationary Operational Environment Satellite
(GOES). The $\rm P_{4}$, $\rm P_{5}$, and $\rm P_{7}$ channel data
we used were 5-minute-averaged and cover the energy ranges 15 --
40, 40--80, and 165 -- 500 MeV, respectively. The
one-minute-averaged cosmic ray intensity was obtained from the
Apatity neutron monitor (NM), which is located at N67.57E33.40. At
the time of the GLE event, Apatity NM was ``viewing" in the
direction towards the Sun along the interplanetary magnetic field
(IMF) lines connecting the Sun with the Earth, so it has a
suitable position for detecting the direct prompt solar cosmic ray
(SCR) intensity.

Assuming that energetic particles travel along the IMF lines at a
speed of $\upsilon$ with no scattering, we usually estimate the
proton solar release time relative to CME white light observation,
by subtracting $\Delta t$ from observed time at 1 AU, where
\begin{equation}\label{1}
    \Delta t=1.1\,\rm AU/\upsilon-8.3\,\rm minutes.
\end{equation}
1.1 AU corresponds to the length of IMF lines when the solar wind
is about 600 km/s for this event. For SCR particles ($\sim$ GeV),
we take $\upsilon$ as approximately 0.9c (c is the velocity of
light). For GOES $\rm P_{4}$ (15 -- 40 MeV) and $\rm P_{7}$ (165
-- 500 MeV)-- considering the velocity dispersion within each
channel and due to the power law spectrum-- the highest
intensities are related to the lowest energies, we take $\upsilon$
as approximately 0.2c and 0.6c, related to the lowest energies 15
and 165 MeV, respectively. Then, the SEP injection profiles as
functions of time and CME-height are plotted in Fig. 4.
\begin{figure}
\centering
\includegraphics[width=8cm]{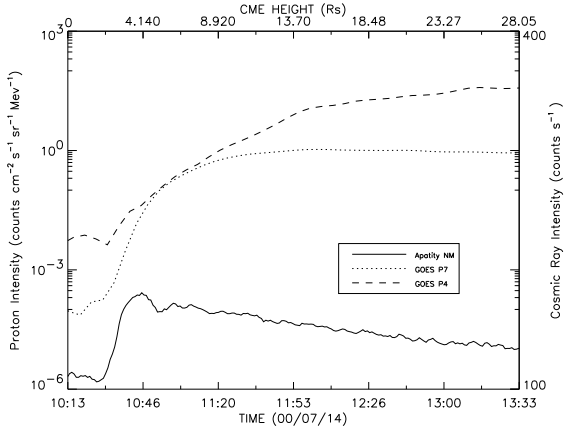}
\caption{SEP injection profiles of the 2000 July 14 GLE. The GOES
$\rm P_{4}$ and $\rm P_{7}$ channel data are 5-minute-averaged and
cover the energy ranges 15 -- 40 and 165 -- 500 MeV, respectively.
Cosmic ray data from Apatity NM is one-minute-averaged.}
\label{fig4}
\end{figure}

The estimate of solar release times in Eq. (1) to determine the
SEP injection profiles is justified only when the SEP mean free
path $\lambda \geq$ 1 AU. However, a range of 0.08$< \lambda <$0.3
AU for SEPs of all energies was found by Palmer (\cite{Palmer82})
from SEP intensity-time and anisotropy-time measurements, so it is
necessary to consider the effect of interplanetary scattering on
the injection profiles of Fig. 4. Using the transport equation
(Kallenrode et al. \cite{Kallenrode92}) and a Reid-Axford
injection profile (Reid \cite{Reid64}), we estimate the
displacements in the plot of Fig. 4 that is delayed from the true
injection profiles. The Reid-Axford injection profile is shown to
be:
\begin{equation}\label{2}
    I(t)=N/t \times \exp(-\beta/t - t/\tau).
\end{equation}
We take ($\beta$, $\tau$, $\lambda$) as (0.1 hr, 0.5 hr, 0.1 AU)
for SCR ($\sim$ GeV) particles and (0.3 hr, 1 hr, 0.08 AU) for
GOES ($\rm P_{4}$ and $\rm P_{7}$) particles, respectively.

Then, the peaks of SCR, $\rm P_{4}$, and $\rm P_{7}$ injection
profiles should be respectively displaced by $\sim$ 20, $\sim$
145, and $\sim$ 78 minutes. Considering these displacements, we
find that the peaks of the injection profiles of SEPs occurs
before 10:31 UT $\pm$ 5 minutes when the CME height reaches $\sim
2 R_{s}$, which is very close to the Sun.

This estimation is consistent with the result of Bieber et al.
(\cite{Bieber02}), who obtained the best fit to the relativistic
protons observed on Bastille Day and found that the injection
function of particles released near the Sun has a peak at 10:26:50
UT with an FWHM (full width half maximum) of 5 minutes. Please
note that this injection function has a good correlation with the
hard X-ray, $\gamma$-ray emission and the inferred reconnection
electric field during 10:22 -- 10:33 UT as Fig. 3 shows.
\begin{figure}
\centering
\includegraphics[width=8cm]{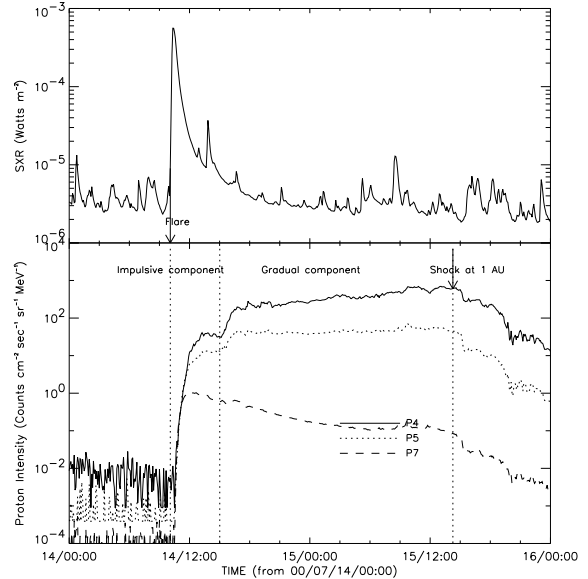}
\caption{Temporal profiles of the soft X-ray and the energetic
proton flux of the 2000 July 14 event. The upper panel shows the 1
-- 8 $\rm\AA$ X-ray integrated flux. In the lower panel, the GOES
$\rm P_{4}$, $\rm P_{5}$ and $\rm P_{7}$ channel data are
5-minute-averaged and cover the energy ranges 15 -- 40, 40--80,
and 165 -- 500 MeV, respectively.} \label{fig5}
\end{figure}

To fully comprehend the SEP event, proton intensity-time profiles
during two days are plotted in Fig. 5, which shows that the
injection profiles have two components or phases, which can be
named ``impulsive component" and ``gradual component". The
impulsive component lasts from 10:26 UT to 15:00 UT on 14 July
2000 only a few hours, followed by the gradual one lasting until
the interplanetary shock reached 1 AU at 14:16 UT on 15 July 2000,
far beyond the flare duration. It is clearly shown that the
gradual component of relative lower energetic protons (P4 and P5)
appears to result from interplanetary shock acceleration, which
however plays a minor role in accelerating somewhat higher
energetic protons (P7).

\section{Discussion}
This section tries to explore the origins of relativistic
particles and the impulsive component of the large gradual SEP
event, flare reconnection acceleration, and shock wave
acceleration?

\subsection{Particle acceleration in the RCS}
In the RCS particles can be accelerated by the DC electric field.
This acceleration process is limited by the gyromotion along the
transverse magnetic field $B_{\bot}$ (Martens \& Young
\cite{Martens90}). The question then arises as to whether the DC
electric field can accelerate protons to GeV energy. Litvinenko \&
Somov (\cite{Litvinenko95}) introduced a transverse electric field
$E_{\bot}$ outside the RCS (Fig. 6), which efficiently ``locks"
the protons in the RCS, thus allowing protons to be accelerated to
a few GeV energy in a time of $<$ 0.1 s. For this event, given the
maximum $E_{rec} \sim 13.0$ V/cm, acceleration length $l_{acc}
\sim 7.0\times10^{7}$ cm is needed to accelerate the protons to
GeV energy. The ratio of the acceleration length to the whole
filament length in this event is $l_{acc}/D \sim 10^{-2}$, hence
protons are not accelerated in a single beam running the full
length of the RCS. This avoids the contradiction that the electric
current associated with the accelerated particles would be so
large that the induced magnetic field would exceed typical coronal
values by a lot.
\begin{figure}
\centering
\includegraphics[width=8cm]{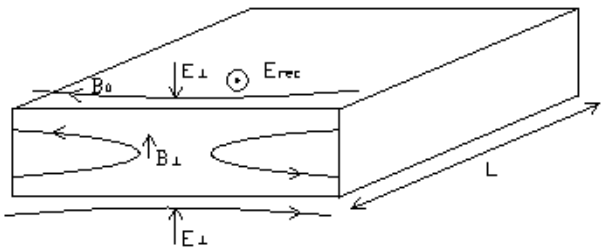}
\caption{Sketch of the reconnection current sheet. $B_{0}$ is the
main (reconnecting) magnetic field component, $B_{\bot}$ the
transverse magnetic field, $E_{rec}$ the electric field related to
the reconnection process inside the sheet, $E_{\bot}$ the
transverse electric field outside the sheet due to electric charge
separation, and L is the length of the current sheet.}
\label{fig6}
\end{figure}

\subsection{RCS -- a reasonable acceleration source of SCR particles}

On the condition described above, protons could be effectively and
rapidly accelerated to a very high energy in the RCS. The inferred
GLE onset time also coincides with the hard X-ray, $\gamma$-ray
emission and the inferred reconnection electric field. It suggests
that the reconnection electric field perhaps makes a crucial
contribution to the acceleration of SCR particles and the
impulsive component of the large gradual SEP event.

An argument against flare acceleration is that the flare did not
occur in the well-connected region. However, Klein et al.
(\cite{Klein01}) find a clear frequency dispersion of the western
radio source during the time around the GLE onset, indicating a
coronal magnetic structure that bends westward with decreasing
frequencies (Fig. 7, left panel). This can be interpreted as a
coronal magnetic channel that links to the well-connected region.
The large-scale solar surface disturbance observed between 10:12
UT -- 10:48 UT is also shown in the right panel Fig. 7. It offers
a particle transport path from the poor-connected region (flare
site) to the well-connected region and may correspond to a
field-opening process.
\begin{figure}
\centering
\includegraphics[width=8cm]{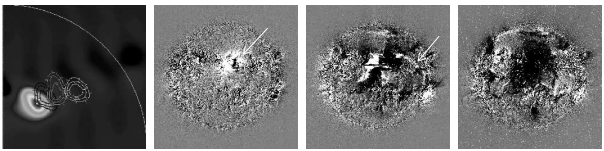}
\caption{Left panel (Klein et al. \cite[Fig. 5]{Klein01}):
Frequency dispersion of the radio source near their maximum
brightness at 410, 327, 237, and 164 MHz. Right panel: The
large-scale solar disk disturbance on 14 July 2000. The three
images show EIT images at 10:12, 10:36, and 10:48 UT with a
pre-event image subtracted from them. Arrow indicates the bright
edge of the solar disturbance.} \label{fig7}
\end{figure}

Another argument against flare acceleration in large gradual
events is that reconnection occurs in closed field lines beneath
CME and that the accelerated particles are trapped so they cannot
escape (Reames \cite{Reames02}). However, according to the time
profiles of hard X-ray and $\gamma$-ray, as well as energetic
particles (including proton, $^{3}$He, and $^{4}$He) from the FY-2
satellite, Tang \& Dai (\cite{Tang03}) found that the 2000 July 14
SEP event should be a mixed one, in which both flare and
CME-driven shock accelerated particles exist. Continuous radio
bursts from a few GHz to $<$ 1 MHz were also reported (Cane et al.
\cite{Cane02}; Klein et al. \cite{Klein01}). It implies that open
field lines extending from within 0.5 solar radii into
interplanetary space must exist, so flare-accelerated particles
can escape along the open field lines into interplanetary space.

\subsection{Other viewpoints}
Indeed, in large SEP events, especially the GLE events, where and
how energetic particles are accelerated remains enigmatic. In the
above section, we suggest that RCS could make a crucial
contribution to the SCR particles and the impulsive component of
the large gradual SEP event. One may propose a possibility that
particles first be energized in the current sheet and then be
reaccelerated by perpendicular coronal shocks (Tylka et al.
\cite{Tylka05}). The large-scale disturbance (Fig. 7, right panel)
may be considered as a visible manifestation of a moving
acceleration region and interpreted as the moving skirt of coronal
shock wave (Cliver et al. \cite{Cliver95}; Torsti et al.
\cite{Torsti99}). However, an important issue is whether
low-energy ions can be efficiently and rapidly accelerated to GeV
energy at the coronal shock given a lower compression ratio (The
compression ratio of this event is $\sim$ 2). Even though
particles can be efficiently and rapidly accelerated by
perpendicular coronal shock, it still cannot explain the two
components of the SEP injection profiles. Tang \& Dai
(\cite{Tang03}) find that the interplanetary shock is the
extension of the coronal shock with the same driver agency --- the
fast CME. As a result, the shock-acceleration SEPs should be
continuously depleting and cannot have two peaks as shown in Fig.
5. Whatever the true cause, the rapid and great enhancement of
relativistic particles cannot be mainly due to shock acceleration,
and the reconnection electric field probably makes a crucial
contribution, while the gradual enhancement of energetic particles
can be perfectly interpreted by CME-driven shock acceleration.

\section{Conclusions}
In this paper, we have evaluated the reconnection electric field
inferred from the 14 July 2000 X5.7 two-ribbon flare, and analyzed
the associated SEP event. Based on the above analysis, our
conclusions can be briefly summarized as follows:

\begin{enumerate}

\item The magnetic reconnection in this event consists of two
parts, corresponding to the evolution of the two parts of the
flare ribbons, respectively. The induced electric field is
temporally correlated with the evolution of hard X-ray and
$\gamma$-ray emission, which indicates that the reconnection
electric field plays an important role in accelerating nonthermal
particles. Given the maximum $E_{rec} \sim 13.0$ V/cm,
acceleration length $l_{acc} \sim 7.0\times10^{7}$ cm is needed to
accelerate protons to GeV energy in the RCS by DC electric field.

\item The injection of SEPs is composed of two components. The
impulsive component occurs very close to the Sun, and the solar
protons' release time is consistent with the hard X-ray,
$\gamma$-ray emission peak and the inferred maximum reconnection
electric field.

\item The reconnection electric field probably makes a crucial
contribution to the acceleration of relativistic particles and to
the impulsive component of the large gradual SEP events, while
CME-driven shocks play a dominant role in the gradual component.
\end{enumerate}

\begin{acknowledgements}
We are very grateful to J. Qiu and the anonymous referee, whose
constructive comments have greatly improved this paper. We thank
the SOHO/MDI, SOHO/EIT, SOHO/LASCO CME catalog, Yohkoh/HXT, and
TRACE teams for providing the observational data. SOHO is a
project of international cooperation between ESA and NASA. Apatity
neutron monitor data is kindly provided by the Polar Geophysical
Institute of the Russian Academy of Sciences (PGI, Apatity). This
work was supported by NKBRSF of China G2000078404 and NSFC key
projects No. 10333040 and No. 10221001.
\end{acknowledgements}

\end{document}